# Space Division Multiplexing in Optical Fibres

D. J. Richardson[1], J. M. Fini[2] and L E. Nelson[3]

[1] *Optoelectronics Research Centre, University of Southampton, Highfield, Southampton, SO17 1BJ, UK.*
[2] *OFS Laboratories,19 Schoolhouse Road, Somerset, New Jersey 08873, USA.*
[3] *AT&T Labs - Research, 200 S. Laurel Avenue, Middletown, New Jersey 07747, USA.*

**Optical communications technology has made enormous and steady progress for several decades, providing the key resource in our increasingly information-driven society and economy. Much of this progress has been in finding innovative ways to increase the data carrying capacity of a single optical fibre. In this search, researchers have explored (and close to maximally exploited) every available degree of freedom, and even commercial systems now utilize multiplexing in time, wavelength, polarization, and phase to speed more information through the fibre infrastructure. Conspicuously, one potentially enormous source of improvement has however been left untapped in these systems: fibres can easily support hundreds of spatial modes, but today's commercial systems (single-mode or multi-mode) make no attempt to use these as parallel channels for independent signals.**

The notion of increasing fibre capacity with Space Division Multiplexing (SDM) is almost as old as fibre communications itself, with the fabrication of fibres containing multiple cores, the first and most obvious approach to SDM, reported as far back as 1979[1]. Yet only recently has serious attention been given to building a complete networking platform as needed to make use of this multicore fibre (MCF) approach. The alternative approach of using modes within a multimode fibre (MMF) as a means to define separate spatially distinct channels also dates back to that era[2].

The current frenzied progress in SDM is occurring now because of a convergence of enabling technological capabilities and a rapidly emerging need. On the one hand, SDM draws on the accumulated progress of fibre research. This includes subtle improvements in traditional fibres[3], and the fantastically precise fabrication methods developed to produce hollow-core and other complex microstructure fibres[4-7]. Sophisticated mode control[8] and analysis[9] methods along with tapered devices[10] can be borrowed from high-power fibre laser research, which itself has needed to develop means to better exploit the spatial domain in the drive to achieve ever higher power levels[11]. Photonic lantern[12] and endoscope devices[13] are available from their development for imaging.

Today's SDM research is also occurring as coherent detection and digital compensation are capable of overcoming complex impairments (such as polarization mode dispersion (PMD)) and are accepted as a standard part of high-performance systems. This is crucial: since SDM packs spatial channels tightly into each fibre, crosstalk between channels is an obvious potential disadvantage and needs to be addressed. The addition of significant crosstalk to a transmission line would have been particularly unattractive a few years ago, before coherent-detection systems offered hope of subtracting out crosstalk electronically at the receiver.

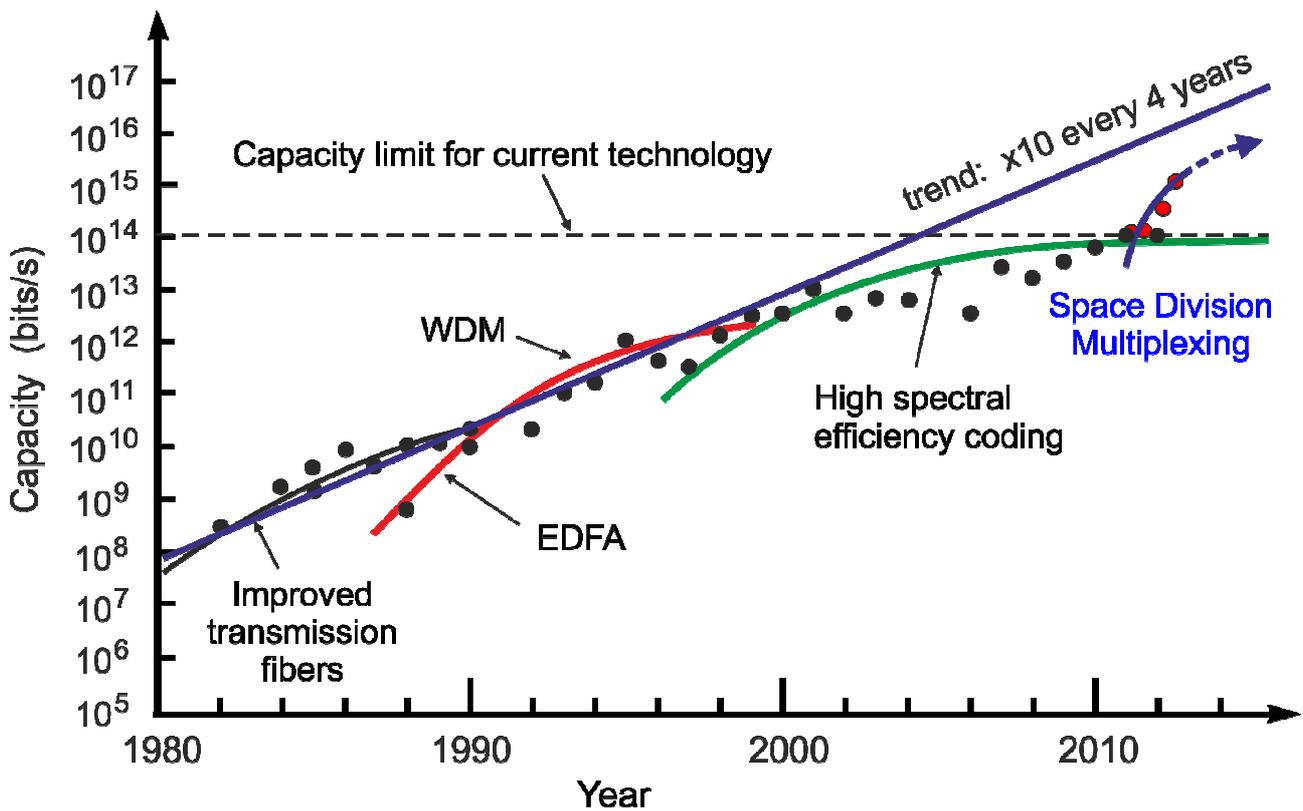

**Figure 1 | The evolution of transmission capacity in optical fibres as evidenced by state of the art laboratory transmission demonstrations over the years**. The data points shown represent the highest capacity transmission numbers (all transmission distances considered) as reported in the Postdeadline Session of the Optical Fiber Communications conference held each year in the USA. The transmission capacity of a single fibre strand is seen to have increased by approximately a factor of 10 every 4 years. Key previous technological breakthroughs include the development of low-loss single-mode fibres, the Erbium Doped Fibre Amplifier (EDFA), Wavelength Division Multiplexing (WDM) and more recently high-spectral efficiency coding via DSP-enabled coherent transmission. The data points for Space Division Multiplexing also include results from the Postdeadline Session at the annual European Conference on Optical Communications (ECOC). As can be seen SDM appears poised to provide the next step change in transmission capacity.

These enabling technologies have made SDM a viable strategy just as a severe need for innovation emerges. Over the past forty years, a series of technological breakthroughs have allowed the capacity-per-fibre to increase around 10x every four years, as illustrated in Figure 1. Transmission technology has therefore thus far been able to keep up with the relentless, exponential growth of capacity demand. The cost of transmitting exponentially more data was also manageable, in large part because more data was transmitted over the same fibre by upgrading equipment at the fibre ends. But in the coming decade or so, an increasing number of fibres in real networks will reach their capacity limit[14]. Keeping up with demand will therefore mean lighting new fibres and installing new cables - potentially also at an exponentially increasing rate. Further, this fibre capacity limit is not specific to a particular modulation format or transponder standard - it is fundamental and can be derived from a straightforward extension of the fundamental Shannon capacity limit to a nonlinear fibre channel under quite broad assumptions[15]. It says that standard single mode fibre (SMF) can carry no more than around 100Tbit/s of data, corresponding to filling the C and L amplification bands of the erbium doped fibre amplifier (EDFA) at a spectral efficiency of ~10 bits/s/Hz.

The upcoming potential "capacity crunch", then, is an era of unfavourable cost scaling. For some carriers who have access to a limited number of dark fibres, very expensive installation of new cables will be the only alternative as the capacity of existing fibres is filled. "Fibre-rich" carriers who attempted to future-proof their fibre plant by including large numbers of premium fibres in each cable (thus putting off the need for subsequent new cables) will be forced to overbuild, i.e. deploy multiple systems over parallel fibres, to keep up with demand. However, multiple systems over parallel fibres suggest that transmission costs and power consumption will scale linearly with growing capacity. The fear is that, without further innovation to lower the cost-per-bit, the capacity crunch will apply pressure to constrain growth, and we will finally reach the end of the seemingly boundless connectivity that drives our economy and enriches our experiences.

The anticipated promise of SDM is not only that it will provide the next leap in capacity-per-fibre, as shown in Figure 1, but that this will concurrently enable large reductions in cost-per-bit and improved energy efficiency[16]. This is a formidable challenge. SDM is very different from wavelength division multiplexing (WDM) which inherently allows the sharing of key components: e.g., an EDFA and dispersion compensation module can easily be shared by many WDM channels with minimal added complexity. The benefits of SDM are more speculative, and assume that many system components can be eventually integrated and engineered to support this potentially disruptive new platform.

Given this emerging need, major research effort has been mobilized around the world to explore and establish the viability of SDM[17]. Exciting recent results show that a wide array of new tools are now being focused on probing the potential benefits of SDM, and chipping away at the many engineering problems obscuring these benefits.

**Technical approaches to SDM**

The term SDM is nowadays taken to refer to multiplexing techniques that establish multiple spatially distinguishable data pathways through the same fibre, although in earlier days the same terminology was previously applied to describe the case of multiple parallel fibre systems: the benchmark that needs to be beaten on a cost-per-bit perspective if any of the SDM approaches currently under investigation are ever to be commercially deployed. The primary technical challenge given the more intimate proximity of the pathways is management of cross-talk.

In the case of multicore fibre (MCF) in which the distinguishable pathways are defined by an array of physically-distinct single-mode cores (Figure 2(b)) the simplest way to limit cross-talk is to keep the fibre cores well-spaced. Small variations in core properties, either deliberately imposed across the fibre cross-section[18], or due to fabrication/cabling[19], can also reduce cross-coupling along the fibre length. As will be discussed later, to date, the highest capacities and longest transmission distances demonstrated in SDM system experiments have all utilized such "uncoupled" MCFs. A study of the tolerance of various advanced modulation format signals to in-band accumulated cross-talk (to include contributions from signal multiplexing and demultiplexing, amplification, splicing and distributed-coupling along the fibre length) showed that < -25dB cross-talk levels are typically required to avoid significant transmission penalties[20].

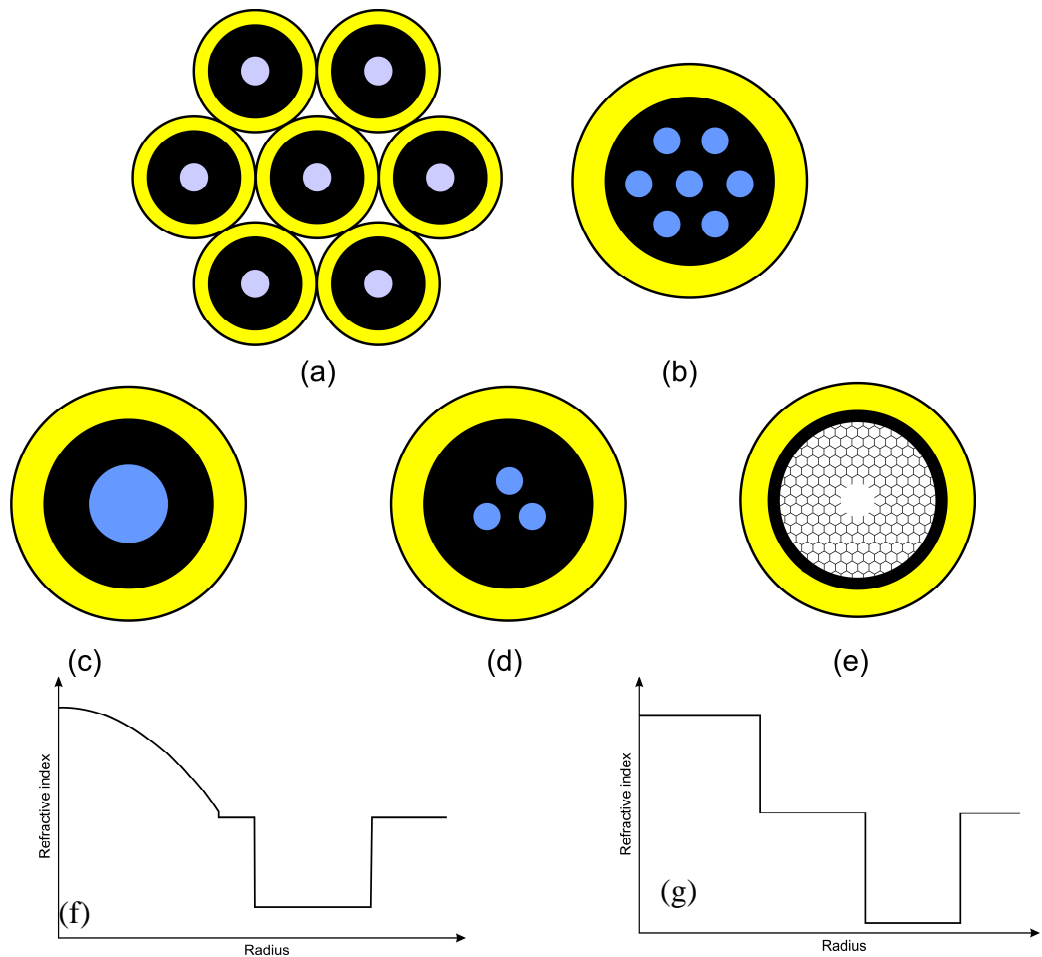

**Figure 2 | Different approaches to SDM:** (a) Fibre-bundles composed of physically-independent, single-mode fibres of reduced cladding dimension could provide for increased core packing densities relative to current fibre cables, however "in-fibre" SDM will be needed to achieve the higher core densities and levels of integration ultimately desired. (b) MCF comprising multiple independent cores sufficiently spaced to limit cross-talk. Fibres with up to 19 cores have so far been demonstrated for long haul transmission – higher core counts are possible for short haul applications (e.g. datacomms) which can tolerate higher levels of cross-talk per unit length. (c) FMF with a core dimension/numerical aperture set to guide a restricted number of modes – so far typically 6-12 distinct modes (including all degeneracies and polarisations). To date work has focussed primarily on using the first few LP-fibre modes; however, work is now beginning on using other modal basis sets that exploit the true vector modes of the fibre – in particular on modes that carry orbital angular momentum and which may provide benefits in terms of reduced mode-coupling and associated DSP requirements[42]. (d) Coupled-core fibres support supermodes that allow for higher spatial mode densities than isolated-core fibres. MIMO processing is essential to address the inherent mode-coupling. (e) Photonic Band Gap fibres[4,5] guide light in an air-core and thus have ultra-low optical nonlinearity, offer the potential for lower losses than solid core fibres (albeit at longer transmission wavelengths around 2μm rather than 1.55μm [43,44]). Work is underway to understand whether such fibres can support MDM and to establish their practicality for high capacity communications. (f) Refractive index profile of a GI core design providing low DMGD and low mode-coupling for long haul FMF transmission[29]. (g) Core refractive index design incorporating a trench profile to reduce cross-talk and thus allow closer core separations in MCF[21].

Using trench-type core refractive index profiles matched to standard SMF (Figure 2(g)), to better confine the mode, it has proved possible to reduce core-to-core coupling to impressively low levels (<-90dB/km) for a spacing of around 40μm, enabling transmission over multi-1000km length scales[21]. However, fibre reliability issues, in particular susceptibility to fracture, mean that MCF diameters beyond 200μm are not considered practical, placing a fairly firm bound on the number of cores that can be incorporated in MCFs for long-haul transmission. Most fibres to date have used a hexagonal arrangement of 7-cores for which the central core, with six nearest-neighbours as opposed to three for cores in the outer ring, suffers the highest level of cross-talk. More recent work[22] has used 12-cores arranged in a ring geometry such that each core has just two nearest-neighbours and experiences nominally the same level of cross-talk (-57 dB/km in this case). A 19-core fibre of 200µm outer diameter has also been reported; however the cross-talk was already substantially higher and limited the useful transmission distance to ~10km[23].

The situation is quite different for mode division multiplexed (MDM) transmission in MMF where the distinguishable pathways have significant spatial overlap and, as a consequence, signals are prone to couple randomly between the modes during propagation. In general the modes will exhibit differential mode group delays (DMGD) and also differential modal loss or gain. The energy of a given data symbol launched into a particular mode spreads out into adjacent symbol time slots as a result of mode-coupling, rapidly compromising successful reception of the information it carries. Crosstalk occurs when light is coupled from one mode to another and remains there upon detection. Inter-symbol interference occurs when the crosstalk is coupled back to the original mode after propagation in a mode with different group velocity. As in wireless systems, equalization utilizing multiple-input multiple-output (MIMO) techniques[24] is required at the receivers to mitigate these linear impairments.

MIMO signal processing is already widely used in current coherent optical transmission systems with polarization division multiplexing (PDM) over standard single-mode fibres. A 2x2 realization with four finite impulse response (FIR) filters recovers the signals on the two polarizations and compensates for PMD in the link[25]. For an MDM system with M modes, the respective algorithms would need to be scaled to 2Mx2M MIMO, requiring $4M^2$ adaptive FIR filters. (By way of comparison, the same capacity carried on M uncoupled SDM waveguides would require 4M adaptive FIR filters.) Thus, if we assume an equal number of taps per adaptive FIR filter and equal complexity of the adaptation algorithm, comparing a 2M×2M MIMO system on M coupled waveguides to M uncoupled SDM waveguides using PDM results in a complexity scaling[26] as $4M^2/(4M) = M$. To compensate DMGD and mode cross-talk completely, the equalization filter length should be larger than the impulse response spread. The computational complexity of FIR filters implemented as time-domain equalizers (TDE) increases linearly with the total DMGD of the link[25], which can make TDE unfeasible for long-haul MDM transmission. Common equalizer algorithms were studied and orthogonal frequency division multiplexing (OFDM) was found to achieve the lowest complexity[27]. However, for OFDM the DMGD to be compensated (and thus the reach) is limited by the length of the cyclic prefix. In other work aimed at lowering the DSP complexity, single-carrier adaptive frequency-domain equalization (SC-FDE) for MDM transmission has recently been proposed[28], where the complexity of SC-FDE scales logarithmically with the total DMGD.

Conventional MMFs with core/cladding diameters of 50/125 and 62.5/125um support more than 100 modes and have large DMGDs, and thus are not suitable for long-haul transmission because the DSP complexity would be too high. Recent advancements have led to fibres supporting a small number of modes, the so-called "few-mode fibres" (FMFs), with low DMGD (see Figure 2(c)). The most significant research demonstrations have so far concentrated on the simplest FMF, which supports three modes, the $LP_{01}$ and degenerate $LP_{11}$ modes, for a total of 6 polarization and spatial modes (referred to as 3MF). The DMGD in step-index core designs (as used in the first demonstrations of MDM in 3MF) is a few ns/km, meaning that the number of taps required for MIMO processing was impractical for transmission distances much greater than 10km. Consequently work has been undertaken to develop core designs offering substantially reduced values of DMGD. Using a graded-index (GI-) core design[29] (Figure 2(f)), DMGD values as low as 50 ps/km have been achieved for 3MF[29,30]. Moreover, it has been shown that DMGD cancellation is possible by combining fibres fabricated to have opposite signs of DMGD[29,31,32]. In this way transmission lines with net values of DMGD as low as ~5ps/km (and with low levels of inherent mode-coupling between mode-groups) have been realised, enabling transmission over >1000km length scales when incorporated with an appropriate amplification approach[33]. Whilst these results are in themselves technically impressive, the question arises as to how scalable the basic approach will ultimately prove. To this end, experiments have been undertaken on both 6-mode[34] and 5-mode FMF[35] with encouraging initial results obtained. However, just as with the MCF approach it is clear that scaling MDM much beyond this is likely to prove very challenging, not least in terms of developing scalable, accurate, low-loss mode launch schemes and ensuring that the required DSP remains tractable. Note that in the future, MCF-based systems may also be designed to use MIMO equalization to deal with increased crosstalk arising from higher core densities and/or highly integrated transmitters/receivers, optical amplifiers, and switching elements.

Whilst zero crosstalk would be ideal, there is a developing school of thought that contends that mode-coupling is inevitable, that full 2Mx2M MIMO is thus necessary, and that strong coupling should be actively exploited [36,37]. If mode-coupling is weak then a data symbol carried by multiple modes with different group indices will spread in time linearly with fibre length. In contrast, if the coupling is strong, then the temporal spread follows a random-walk process, and will scale with the square-root of fibre length. Strong coupling can therefore potentially reduce the number of MIMO taps required and consequently the DSP complexity. Indeed this is analogous to spinning of current single-mode fibre during fabrication to reduce PMD. Similarly, the impact of differential modal gain and loss can in principle be mitigated by strong mode-coupling over a suitable length scale relative to the amplifier spacing[38,39]. In the MCF case, by bringing the cores closer together to ensure strong mode-coupling, it is possible to establish supermodes defined by the array of cores, which can then be used to provide spatial information channels for MDM to which MIMO can be applied[40,41]. This enables higher spatial channel densities for MCFs than can be obtained using isolated cores designs.

**SDM Technology and Integration**

While development of innovative fibres for SDM goes on, researchers have turned to component and connectivity challenges that are essential to building systems around SDM fibres. Recent component demonstrations give us glimpses of what is possible, but are based on varied and often conflicting assumptions about the larger system. Two visions of a total system are of particular interest: the "grand vision" of an ultra-high capacity, fully-SDM system, and the "upgrade-path" vision, where SDM components and links are gradually added to existing non-SDM infrastructure.

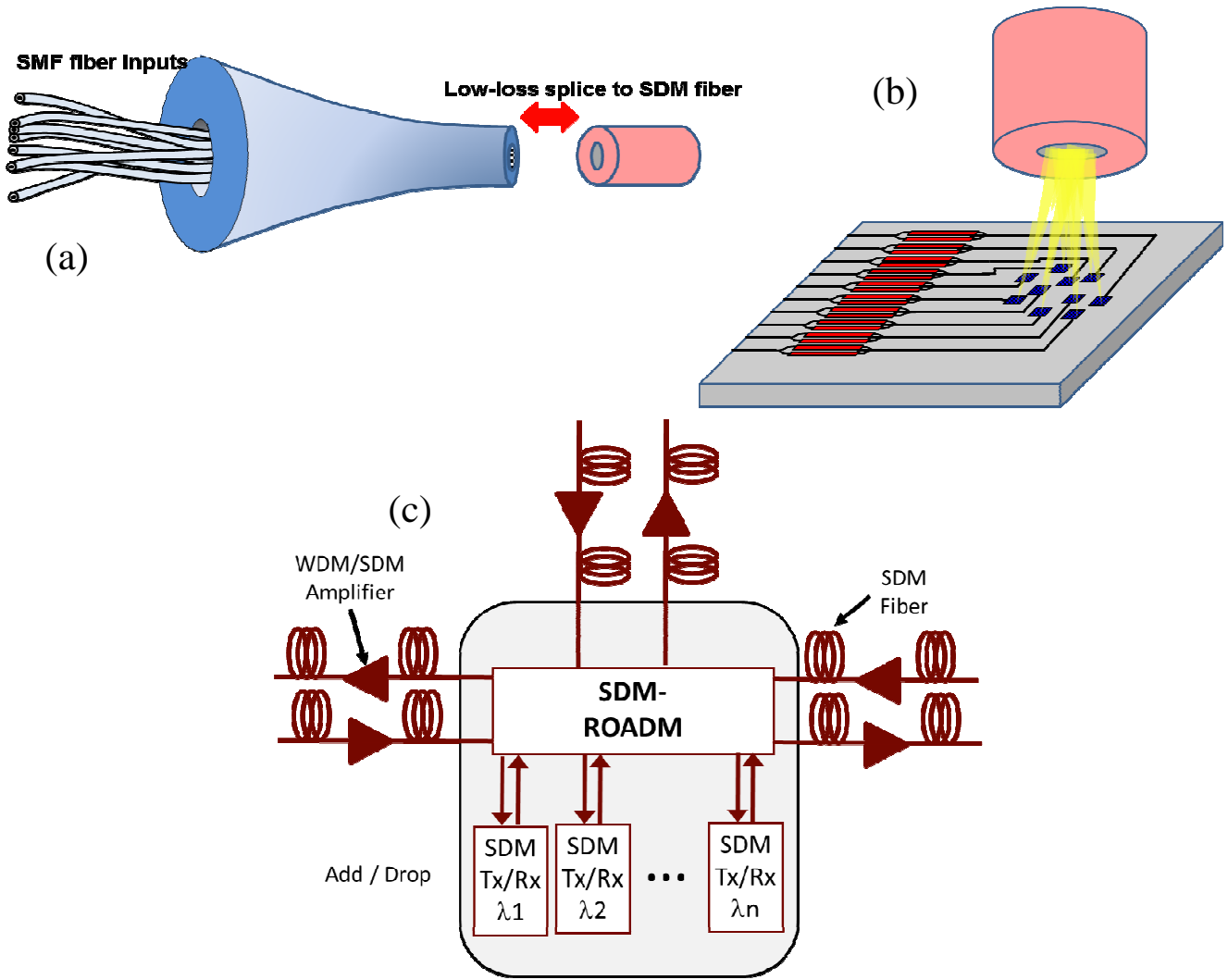

**Figure 3 | Many components are needed to fully utilize the benefit of high-density SDM, including** (a) elegantly scalable passive multiplexers, (b) integrated transmitter and receiver arrays providing low-loss coupling to many modes of an SDM fibre, and (c) reconfigurable routing elements that can direct SDM traffic without the need for electronic MIMO in between transmitter and receiver. The strategy for switching wavelengths and modes, as well as any additional required functionality, will determine the complexity of the ROADM architecture.

When building components for the fully-SDM vision, we look into an eventual future where high performance MIMO is available, and where the density of spatial multiplexing must be increased well beyond the low-crosstalk limits. Scalability takes primary importance, whereas low-crosstalk designs confer no benefit. These broad guidelines suggest what component types will be preferred: the constituent modulators or detectors in a transmitter or receiver should be coupled to an SDM fibre in a very scalable way without regard for any one-to-one mapping of signals to modes (since the modes will quickly scramble in any case). Such scalability is natural, for example, in photonic-lantern multiplexers[12,45,46] illustrated in Figure 3(a) and in spot-type transmitters[47] illustrated in Figure 3(b). A real system will use passive multiplexers wherever possible, to avoid cost and power dissipation. Flexible devices for actively multiplexing a desired spatial channel, such as those based on spatial light modulators[48,49], may be important, but only in the small number of subsystems where reconfigurable optical add/drop functions for spatial channels are needed.

Amplifier design is somewhat less clear in this fully-SDM regime. The potential cost and power savings of a cladding-pumped architecture are quite attractive[50]. On the other hand, demonstrations of few-mode core-pumped amplification with low differential gain are interesting[51,52] and may perhaps be scaled to larger number of modes. In either case, further work will be needed to develop gain fibres that can achieve high efficiency and low differential gain for many modes.

Today's flexible photonic mesh networks are based on reconfigurable add-drop multiplexers (ROADM), which provide carriers the ability to remotely establish lightpaths and efficiently switch those lightpaths on demand. Similar routing flexibility is assumed in future fully-SDM networks; however, there are many options for how this could be attempted. Figure 3(c) shows an example of a possible SDM network node, consisting of an SDM ROADM and associated SDM-WDM transmitters and receivers for the add and drop, along with SDM fibres and amplifiers for the three fibre directions addressed. The SDM ROADM itself could be quite complicated depending on the choices made regarding the type of superchannels (e.g. frequency or spatial[53]) and the strategy for switching, for example: 1) wavelength by wavelength, all modes together; 2) mode by mode, with mode interchange; 3) any wavelength/mode to any other wavelength/mode. The SDM-ROADM architecture also will depend on its support for colorless, non-directional, contentionless, and/or gridless operation.

The general strategy that mode-coupling is inevitable and should be sorted out only at the receiver suggests that mode-independent routing elements will be a key technology: these route wavelengths to different destinations while keeping all spatial modes together. All modes will then be present at the receiver, allowing MIMO processing to recover the transmitted signals. A first demonstration of a few-mode-compatible OADM was reported by Chen et al.[54]. Spatial superchannels and the plausibility of joint digital signal processing have also been considered[53]. Cvijetic et al.[55] give a forward-looking view of SDM as a tool for ultimate flexibility in routing. As it matures, SDM may offer benefits beyond high capacity and flexible routing, for example in secure data transmission[56].

In a very different vision of SDM, upgrades of an existing single-mode infrastructure are incorporated incrementally. They must be reverse-compatible and offer cost advantages in the short term. The key here is not scalability, but compatibility, and so low-crosstalk solutions are extremely important: they allow SDM fibres and components to be used as drop-in replacements for their non-SDM counterparts (leading to a "hybrid SDM-SMF" network), and need not wait until real-time implementations of MIMO processing have been commercialized. This vision does not contradict the assumptions of the "grand vision," so much as it looks at a different time scale. Initial steps down an SDM "upgrade path" could occur quite soon, including, for example, replacing cables in a few ducts of unusually high congestion, without changing the surrounding network. Seamless connectivity will be a key requirement in early upgrades. Basic passive multiplexers have been demonstrated using free-space[57] and tapered[58] approaches and both are already rapidly progressing towards more practical and compact solutions [22,59,60].

Upgrading a conventional SMF system with SDM amplifiers could offer considerable advantages if multicore amplifiers achieve the efficiency improvements potentially available[50,61] from cladding-pumping and pump sharing. This would provide system companies an incremental upgrade with a clear motivation, but is technically challenging. Initial steps towards a cladding-pumped MCF-EDFA[62,63] are interesting, but are far from achieving high efficiency.

A key motivation for SDM is its potential to facilitate integration, and this will be a priority both in incremental upgrades and later progress towards fully SDM systems. Synergistic development of integrated transmitter and receiver arrays along with compatible fibres and components is essential, and was first seen in short-reach data communications systems[64,65]. Silicon integrated devices matched to fibre with multiple single-mode cores have been demonstrated with seven cores[66] and eight cores[67]. Devices for producing the complex orthogonal fields matching specific modes of a MMF have also been demonstrated[68-70]. Many other functions are needed in a real system. For example, a Raman pump sharing circuit[71] suggests how component counts in transmission equipment can be greatly reduced by integration. The net benefit of integration will be clearer as engineering of interconnects and multiplexers brings total loss in line with single-mode devices.

**Progress in Systems Demonstrations**

Initial transmission experiments over MCF were aimed at short-reach applications. Following a proposal to make high density cables for FTTH applications[72], simultaneous 850nm, 1Gb/s transmission over two cores of a four-core MCF was first demonstrated by Rosinski et al.[73]. More recently, 1310nm and 1490nm signals were transmitted over an 11.3-km seven-core MCF for passive optical network applications, where a fibre-based tapered multicore connector (TMC) was first utilized to couple signals into and out of the MCF[74]. For optical data link applications, seven-core multimode MCFs fabricated from graded-index core-rods were also demonstrated in 2010, with transmission 7 x 10-Gb/s over 100m using TMCs and discrete 850nm VCSELs[75]. Lee et al.[54] reported a hexagonal array of VCSELs for coupling directly to the outer six cores of the seven-core multimode MCF, and subsequently developed a vertically illuminated photodiode array matched to the MCF[66], thus demonstrating a full 100m MCF optical link at up to 120Gb/s

Following significant efforts on the design and fabrication of single-mode MCFs, demonstrations of SDM transmission over MCF for long-haul applications have shown impressive progress in terms of capacity, reach, and spectral efficiency, as detailed in Table 1. The first WDM transmission experiments over MCFs were simultaneously reported by two groups using seven-core MCF, with 56Tb/s capacity over 76.8km[76] and 109Tb/s capacity over 16.8km[77]. In both experiments, the MCF cross-talk was sufficiently low such that the signals on each core could be received independently (without MIMO processing), and optical amplification utilized conventional single-core EDFAs placed before and after the MCF. Over the past two years, a number of subsequent experiments over seven-core MCFs have been performed with spectral efficiencies of 14 bit/s/Hz or higher[78,79].

| Year | Reference | Fiber Type | Number of cores/ modes | Distance (km) | Span length (km) | Channel Rate (Gb/s) | WDM channels in each core/mode | Net Spectral Efficiency (b/s/Hz) | Net Total Capacity (Tb/s) | Capacity-Distance Product (Pb/s * km) |
|---|---|---|---|---|---|---|---|---|---|---|
| 2012 | 22 | MCF | 12 | 52 | 52 | 456 | 222 | 91.40 | 1012.32 | 52.64 |
| 2012 | 82 | MCF | 7 | 6160 | 55 | 128 | 40 | 14.44 | 28.88 | 177.87 |
| 2012 | 81 | MCF | 7 | 845 | 76.8 | 603 | 8 | 42.20 | 33.77 | 28.53 |
| 2012 | 23 | MCF | 19 | 10.1 | 10.1 | 172 | 100 | 30.50 | 305 | 3.08 |
| 2012 | 96 | µstr-MCF | 3 | 4200 | 60 | 80 | 5 | 3.84 | 0.96 | 4.03 |
| 2011 | 79 | MCF | 7 | 2688 | 76.8 | 128 | 10 | 15.02 | 7.51 | 20.19 |
| 2011 | 80 | MCF | 7 | 76.8 | 76.8 | 1120 | 1 | | 7.84 | 0.60 |
| 2011 | 41 | µstr-MCF | 3 | 1200 | 60 | 80 | 1 | | 0.22 | 0.27 |
| 2011 | 40 | coupl MCF | 3 | 24 | 24 | 56 | 1 | | 0.15 | 0.004 |
| 2011 | 58 | MCF | 7 | 76.8 | 76.8 | 107 | 160 | 14.00 | 112.00 | 8.60 |
| 2011 | 76 | MCF | 7 | 76.8 | 76.8 | 107 | 80 | 14.00 | 56.00 | 4.30 |
| 2011 | 77 | MCF | 7 | 16.8 | 16.8 | 172 | 97 | 11.25 | 109.14 | 1.83 |
| 2012 | 99 | MCF with SM and FM cores | 12 SM cores, 2 FM cores | 3 | 3 | 1050 | 385 in SM cores, 354 in FM cores | 109.00 | 1050.00 | 3.15 |
| 2012 | 34 | FMF | 6 | 130 | 65 | 80 | 8 | 7.68 | 3.07 | 0.40 |
| 2012 | 32 | FMF | 3 | 119 | 119 | 256 | 96 | 12.00 | 57.60 | 6.85 |
| 2012 | 94 | FMF | 3 | 209 | 209 | 80 | 5 | 4.42 | 1.10 | 0.23 |
| 2012 | 33 | FMF | 3 | 1200 | 30 | 80 | 1 | | 0.19 | 0.23 |
| 2012 | 93 | FMF | 3 | 85 | 85 | 112 | 1 | | 0.31 | 0.03 |
| 2012 | 91 | FMF | 3 | 96 | 96 | 80 | 1 | | 0.22 | 0.02 |
| 2011 | 35 | FMF | 5 | 40 | 40 | 112 | 1 | | 0.52 | 0.02 |
| 2011 | 92 | FMF | 3 | 50 | 50 | 112 | 88 | 6.19 | 27.23 | 1.36 |
| 2011 | 33 | FMF | 3 | 33 | 33 | 112 | 6 | 6.19 | 1.86 | 0.06 |
| 2011 | 90 | FMF | 3 | 10 | 10 | 56 | 1 | | 0.15 | 0.002 |
| 2011 | 89 | FMF | 2 | 40 | 40 | 112 | 1 | | 0.21 | 0.01 |
| 2011 | 88 | FMF | 2 | 4.5 | 4.5 | 107 | 1 | 5.40 | 0.20 | 0.001 |

**Table 1| Summary of progress in SDM system experiments.** Upper rows (shaded in purple) show experiments utilizing multicore fibres (MCF). Center row (shaded blue) shows a MCF-FMF result. Lower rows (shaded in pink) show transmission results over few-mode fibres (FMF). The channel rate includes polarization multiplexing; the Net Spectral Efficiency and Net Total Capacity exclude the overhead for forward-error-correction. SM: single-mode, FM: few-mode, coupled MCF: coupled-core MCF, µstr-MCF: microstructured, coupled-core MCF. The majority of SDM transmission experiments have utilized either 7-core MCF or FMF supporting 3 spatial modes. The table shows the rapid progress in both reach and net capacity achieved in just two years. Recent experiments utilizing a 12-core MCF and a MCF with hybrid SM and FM cores have demonstrated record net capacities of more than 1 Pb/s.

Several experiments have utilized coherent optical orthogonal frequency-division multiplexed (CO-OFDM) superchannels to demonstrate ultra-high per-channel bit rates over MCF. Liu and co-workers transmitted a single 1.12Tb/s superchannel comprised of 20 OFDM subchannels over a single 76.8km span of MCF[80]. A later WDM experiment demonstrated transmission of eight 603Gb/s superchannels over 845 km and achieved 42.2 bit/s/Hz spectral efficiency[81].

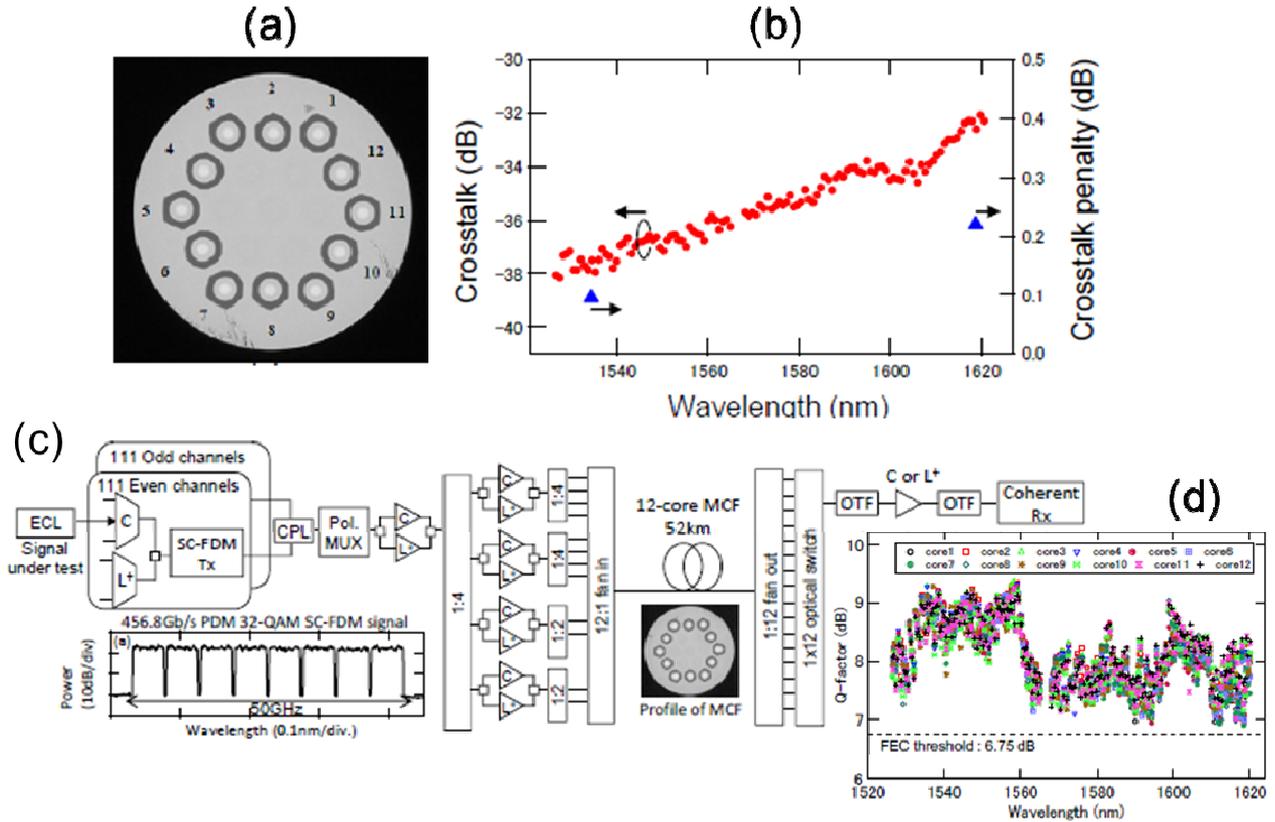

**Figure 4 | 1.01 Pbit/s MCF WDM/SDM/PDM transmission experiment[22].** (a) Microscope image of the cross section of the one-ring, 12-core fibre. (b) Total crosstalk from all other cores after transmission over the 52km 12-core MCF and fan-in/fan-out devices (filled circles), and crosstalk penalty measured for two channels (filled triangles). (c) Schematic diagram of the transmission system setup. ECL: external cavity laser, SC-FDM Tx: single-carrier frequency-division multiplexed transmitter, CPL: coupler, Pol Mux: polarization multiplexer, OTF: optical tunable filter, Rx: receiver. (d) Measured Q-factors of the 222 WDM channels in each of the 12 cores after 52-km transmission.

The longest transmission distance over MCF was recently reported by Takahashi and co-workers[82], who transmitted forty 103Gb/s channels over 6160 km. That experiment was the first to utilize a multi-core EDFA for long-haul, WDM transmission over MCF, and a record capacity-distance product of 177 (Pb/s)*km was achieved. A record capacity of 1.01Pb/s was transmitted over a 52km MCF with 12 cores arranged in a single ring[22], as shown in Figure 4(a). This MCF and its fan-in and fan-out devices were specifically designed for sufficiently low cross-talk to allow high-order modulation formats[20], in this case 32-QAM at 400Gb/s per channel. In addition to supporting an aggregate spectral efficiency of 91.4 bit/s/Hz, the MCF had sufficiently low loss and crosstalk across the C- and extended L-bands (1526 to 1620nm), as displayed in Figure 4(b), thus enabling transmission of 222 channels to achieve the 1.01 Pb/s capacity (Figures 4(c,d)).

The MDM concept was first proposed in 1982 when Berdague and Facq used spatial filtering techniques to launch and detect two modes at the ends of a 10m conventional graded-index MMF[2]. It was not until 2000 when Stuart recognized the analogy between wireless and optical channels and demonstrated the application of 2x2 MIMO for reception of two MDM channels after 1km transmission[83]. Coherent optical 2x2 MIMO was demonstrated at 800Mb/s over 100m and 2.8km of 62.5-μm MMF[84]. Additional proof-of-principle demonstrations utilized direct-detection and 2x2 MIMO[85], and modal-diversity with direct-detection using 2x4 MIMO[86]. These early transmission experiments all over conventional MMF did not launch and receive all modes, and thus would not have been capable of achieving the low-outage expected of optical communication links[87].

Recent FMF development has enabled rapid progress in the capacity and reach of MDM systems demonstrations, as shown in Table 1. In early 2011, three experiments over the simplest FMF supporting the $LP_{01}$ and degenerate $LP_{11}$ modes were reported at the same conference. Per-channel rates of 100Gb/s over two spatial modes were achieved over 4.5km[88] and 40km[89], while 56Gb/s signals in three modes were transmitted over 10km[90] (the quoted bit-rates include polarization multiplexing). The three-mode experiment[90,91] first demonstrated full use of all degrees of freedom afforded by the FMF (6 spatial and polarization modes), with signal recovery via full coherent 6x6 MIMO. Randel et al[26] subsequently demonstrated WDM, MDM and PDM transmission of 6 wavelengths, 3 modes, and 2 polarizations, achieving 2.02Tb/s capacity (before subtracting the FEC overhead).

In the first transmission experiment to include amplification in a MMF, a few-mode inline EDFA boosted the 88 WDM signals before the mode demultiplexer and reception[92]. To reduce mode-dependent gain, the MM-EDFA was forward pumped in the $LP_{21}$ mode and reverse pumped in the $LP_{11}$ mode. In later work, the transmission distance was increased to 85km[93]. Distributed Raman amplification was employed to demonstrate single-span MDM transmission over 209km[94]. In that experiment, the FMF supporting 6 spatial and polarization modes consisted of a first section of large effective area (>155um$^2$) depressed-cladding FMF (which is tolerant to nonlinear effects) followed by a graded-index FMF with smaller effective area (<67um$^2$) for efficient Raman pumping. The DMGD was compensated by using fibres spools with DMGD of opposite sign to reduce the required number of equalizer taps in the MIMO processing.

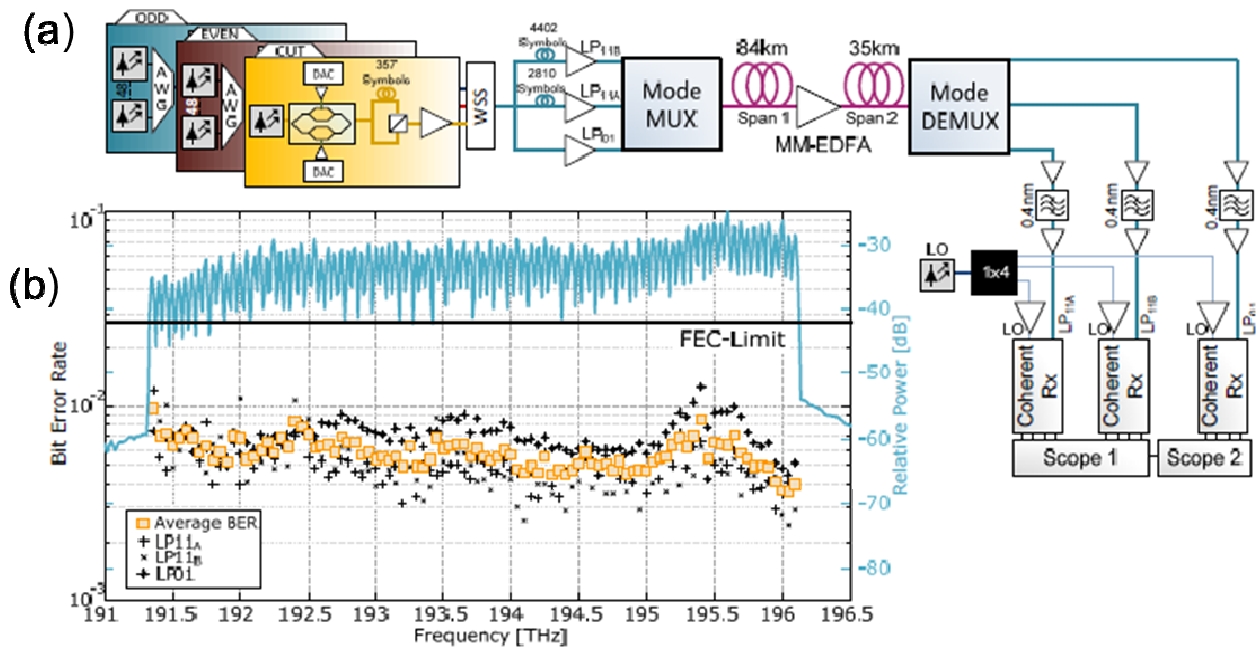

**Figure 5 | 57.7 Tbit/s amplified WDM/MDM/PDM transmission experiment over a few-mode fibre[32].** (a) Schematic of the experimental setup, showing the mode multiplexer and demultiplexer and simultaneous reception of the channels transmitted in the three modes for MIMO processing. CUT: channel-under-test, AWG: arrayed waveguide grating multipler, DAC: digital-to-analog converter, WSS: wavelength-selective switch, LO: local oscillator. (b) Measured bit-error-rates (markers) and optical spectrum of all 96 channels after transmission over the 119km of few-mode-fibre with a mid-span amplifier.

To date, the longest transmission distance reported for MDM systems is 1200km, where a recirculating loop was utilized with a 30km span consisting of two FMF sections having DMGDs of opposite sign[33]. Mode multiplexers and demultiplexers were placed before and after the FMF span in the loop so that single-mode EDFAs could be used. Recently, a net capacity of more than 57Tb/s (after subtracting the overhead for FEC) was demonstrated in a 119km MDM system (Figure 5(a)) consisting of 96 WDM channels each at 200Gb/s[32]. The FMF supported three spatial modes, and an inline MM-EDFA provided 18dB of gain per mode, making this the first WDM FMF system to utilize a mid-span MMF amplifier. The 12bit/s/Hz spectral efficiency and 57Tb/s capacity are the highest reported so far for MDM transmission. The highest number of spatial modes to be utilized as separate information channels was reported by Ryf and co-workers[34], who used six spatial modes ($LP_{01}$, $LP_{11}$, $LP_{21}$, and $LP_{02}$) and 12 x 12 MIMO.

**Combining MCF and FMF concepts**

As shown in Figure 2(d) multicore fibres with coupled cores[95] allow increased core density and/or the cores' effective areas to be increased to minimize nonlinear effects. Due to the strong cross-talk between cores, the light propagation in the fibre can be described by the supermodes of the composite fibre structure. A single-channel transmission experiment with a 24-km homogenous three-core fibre with 104um$^2$ effective areas was reported[40], where all six (space and polarization) modes were launched and jointly detected. The large crosstalk of about -4dB was almost completely suppressed by coherent 6x6 MIMO processing. In other work, an all-solid-glass microstructured fibre with three large (~129um$^2$) effective area cores at 29.4um pitch was utilized in several experiments[41,95]. Using 6×6 MIMO processing, transmission over 1200km was achieved for a single 20-Gbaud-QPSK channel and over 4200km for five WDM channels[96], a record distance at that time, thus demonstrating the feasibility of MIMO interference cancellation for long-haul transmission.

In order to gain a more significant capacity increase it is possible to combine the MC and MDM approaches, and indeed the first experiments in this area are now beginning. For example, it is possible to produce an array of spatially isolated MM rather than SM cores[97,98], enabling MDM to be overlaid directly on to the MC approach. To date fibres capable of supporting up to 21 different spatial modes (7 cores each supporting 3 modes) have been reported although, as of yet, only relatively rudimentary system measurements have been performed[97]. In a related approach, a hybrid MCF with 12 single-mode cores and two few-mode cores supporting three spatial modes each has been utilized to demonstrate transmission of 1.05 Pb/s capacity[99]. Although the transmission distance was only 3km, the experiment was the first to achieve spectral efficiency beyond 100 bit/s/Hz. These approaches are extremely challenging from a component perspective, nevertheless they provide interesting opportunities both in terms of constraining DSP complexity and allowing very high spatial channel densities

**SDM Networking and Switching**

Although to date the majority of the SDM demonstrations have consisted of point-to-point transmission, recent efforts are contemplating switching strategies and elements that could support flexible optical routing. Consensus is building that SDM networks should utilize spatial superchannels (i.e. groups of same-wavelength subchannels transmitted on separate spatial modes but routed together, where the spatial modes could be the regular modes in a MMF/FMF, super-modes in a strongly coupled multi-core fibre, or the fundamental modes of each individual single-mode core in an 'uncoupled' multi-core fibre)[53,54]. Such a strategy could provide sufficient granularity for efficient routing and facilitate ROADM integration, and could help to simplify network design since the modes are routed as one entity, foster transceiver integration (e.g. share a single source laser in the transmitter and a single local oscillator in the receiver), and lighten the DSP load by exploiting information about common-mode impairments such as dispersion and phase fluctuations[53].

As a first-step towards a FMF-compatible ROADM, an optical add-drop multiplexer (OADM) comprising two cascaded, free-space, thin-film filters has been demonstrated for the two orthogonal $LP_{11}$ modes[54]. Recently, Amaya et al.[100] reported switching in the space, frequency and time dimensions in an elastic SDM and multi-granular network that included two 7-core MCF links. Space switching was achieved via an optical backplane that interconnected MCF/SMF fibre inputs, functional modules, and MCF/SMF fibre outputs; however, there was a high degree of complexity due to the required demultiplexing of the signals transmitted on the various cores MCF at each node. Future work is needed to examine the trade-offs between the switching granularity and resulting complexity in SDM networks.

**Conclusions**

Over the past two years much exciting progress has been made in SDM transmission. Various technological approaches are under development and record per-fibre capacities and long reaches have already been demonstrated, some including SDM amplifiers. Initial demonstrations of switching/routing have also now been performed. However this is just the beginning and much further work needs to be undertaken if per-channel reliability and performance competitive with existing single-mode links is to be achieved. Further, most network operators will only consider deploying SDM if it (1) lowers the cost-per-bit, (2) provides the routing flexibility needed for efficient photonic mesh networks, and (3) allows a reasonable transitional strategy from systems based on standard SMF. Photonic integration will be absolutely essential, and this work is just at its infancy. Whatever the ultimate outcome, the next few years promise to be a busy and exciting time in optical fibre communications research.